\documentclass[aps,pra,floatfix,twocolumn]{revtex4-2}

\usepackage{longtable}
\usepackage{dcolumn}
\usepackage{bm}
\usepackage{bbm}
\usepackage{slashed}
\usepackage{times}
\usepackage{nicefrac}
\usepackage{amsmath}
\usepackage{amsfonts}
\usepackage{amssymb}
\usepackage{amsthm}

\newcolumntype{.}{D{x}{}{-1}}
\newcolumntype{w}[1]{D{.}{.}{#1}}
\newcommand{\Za}{Z\alpha}

\allowdisplaybreaks

\begin{document}

\title{Radiative corrections to the nuclear size and polarizability effects in atomic systems}

\author{Krzysztof Pachucki}
\affiliation{Faculty of Physics, University of Warsaw,
             Pasteura 5, 02-093 Warsaw, Poland}

\begin{abstract}
We perform a complete calculation of $\alpha\,(Z\,\alpha)^5\,m$ radiative corrections 
to the finite nuclear size, the recoil finite size and the nuclear polarizability effects in atomic systems.
Results confirm very good agreement for the mean square charge radii difference $r_d^2-r_p^2$
between the deuteron and the proton, as measured in {\em electronic} and muonic isotope shifts.
\end{abstract}

\maketitle

\section{Introduction}

With the increased precision of atomic isotope shift measurements \cite{amsterdam:25} 
and muonic atom determinations of the nuclear charge radii \cite{crema:25},
knowledge of nuclear recoil, finite nuclear size (fns), and polarizability effects became of great interest. 
The accurate calculation of these effects is important for the comparison of {\em electronic} and muonic atom isotope shift, 
from which one obtains the difference in the mean square charge radii between isotopes \cite{pachucki:24a}. 
This difference is an important test of  the low-energy interactions.
For example, the second-order recoil correction $(Z\,\alpha)^6\,m^3/M^2$
was recently found \cite{qi:25, pachucki:24b} to be significant for the  $^3$He-$^4$He isotope shift,
due to the hyperfine mixing, and resolved a discrepancy between the {\em electronic}
and muonic determination of the mean square nuclear charge radii difference.

The nonradiative fns corrections including the leading recoil $(Z\,\alpha)^6\,m^2/M$ have been the subject of recent works 
\cite{pachucki:23, yerokhin:23, pachucki:25a}.  The exact result without expansion in $Z\,\alpha$ in the nonrecoil limit
is a matter of solving the Dirac equation with extended size nucleus. It is nevertheless valuable to perform
the $Z\,\alpha$ expansion, because the explicit dependence on the charge distribution can be obtained.
For example, at the order of $(Z\,\alpha)^6\,m$ the finite size effect includes the $r^2_C\,\ln (m\,r_C)$ term.
Leading recoil corrections can also be calculated exactly 
using the recently derived formula for the recoil correction with extended size nuclei \cite{pachucki:23, yerokhin:23},
while the $Z\,\alpha$ expansion leads to the $r_C^2\,\ln(m\,r_C)$ term at $(Z\,\alpha)^5\,m^2/M$ and a linear $r_C$ term at  $(Z\,\alpha)^6\,m^2/M$ orders.

In contrast to the above, the exact calculation of radiative finite size corrections is much more difficult, because it involves 
the electron self-energy. They have been calculated numerically in Ref. \cite{yerokhin:11},
but it is difficult to obtain the functional dependence  on the nuclear charge distribution  in this way. 
Therefore, they have been calculated analytically in the leading order $\alpha\,(Z\,\alpha)^5$ in Refs. \cite{pachucki:93a, eides:01, sgk:18}
and at some higher orders \cite{yerokhin:11, sgk:18}.
It is even more difficult to obtain the leading radiative recoil correction to the finite size effect.
The exact formulas have recently been derived using  a formulation of {\sl Heavy Particle QED} (HPQED)\cite{pachucki:24c}, 
but they have not yet been implemented numerically. 
In fact, the radiative recoil correction is the main source of uncertainty in hydrogen energy levels \cite{codata:22},
after the improved numerical calculation of the two-loop electron self-energy has been performed \cite{yerokhin:24}.
 
In this work, we collect all known results for radiative finite size corrections at leading orders,  
present their comprehensive derivation, and in particular,
obtain  finite nuclear size effects of order $\alpha\,(Z\,\alpha)^5\,m$ and $\alpha\,(Z\,\alpha)^5\,m^2/M$,
which are important for determination of the nuclear charge radii from the {\em electronic} and muonic atomic spectroscopy.
In addition we perform the calculation of the electron self-energy correction
to the nuclear polarizability effect, which is currently the main source of uncertainty in H-D isotope shift~\cite{pachucki:18}.

Although calculations are presented for hydrogenic systems, results are valid for arbitrary (few-electron) light atoms and ions,
because they are all proportional to $\phi^2(0) = \langle\phi|\sum_a\delta^3(r_a)|\phi\rangle$ at $\alpha(Z\,\alpha)^5$ order.

\section{Expansion of the finite nuclear size effect in powers of $\alpha$}
Let us denote by $E_{\rm fns}$ the shift in the binding energy of a hydrogenic system due to the finite nuclear size (fns).
For a light atom we can perform the expansion of $E_{\rm fns}$ in the small nuclear charge $Z\,\alpha$ and in $\alpha$ (together)
\begin{align}
 E_{\rm fns} =  E^{(4)}_{\rm fns} +  E^{(5)}_{\rm fns}+  E^{(6)}_{\rm fns}  + \ldots, \label{01}
 \end{align}
where the superscript indicates the order in $\alpha$.
The leading-order nuclear contribution is of order $(\Za)^4$ and is given by a simple formula,
\begin{equation}
  E^{(4)}_{\rm fns} =  \frac{2\,\pi}{3}\,\Za\,\phi^2(0)\,r_C^2\,, \label{02}
\end{equation}
where $\phi(0)$ is the nonrelativistic wave function of the electron at the position of nucleus,
$r_C$ is the root-mean-square charge radius of the nucleus
\begin{align}
r_C^2 = \int d^3 r\, r^2\,\rho(\vec r), \label{03}
\end{align}
and $\rho(\vec r)$ is the nuclear charge distribution. Equation (\ref{02}) includes the exact dependence on
the finite nuclear mass $M$
through $\phi^2(0) = \mu^3\,(\Za)^3/(\pi n^3)$, where $\mu = mM/(m+M)$.
The mass dependence of higher-order corrections is much more complicated,
and it is convenient to perform an additional expansion in the mass ratio, namely
\begin{align}
E^{(n)}_\mathrm{fns} = E^{(n,0)}_\mathrm{fns}  + E^{(n,1)}_\mathrm{fns}  + \ldots, \label{04}
\end{align}
where the second superscript denotes a power in the $m/M$ expansion coefficient. 
The leading expansion terms are known, but for the clarity of presentation 
we will repeat their derivation here and start from 
$E^{(5,0)}_\mathrm{fns}$.

\section{$(Z\,\alpha)^5$ Nonrecoil  \MakeLowercase{FNS}}
The next to leading order $E^{(5,0)}_{\rm fns}$ finite size correction in the nonrecoil limit
is given by the two-photon exchange amplitude $T^{\mu\nu}$
\begin{align}
E^{(5,0)}_\mathrm{fs} =&\ \phi^2(0)\,\int \frac{d^3q}{(2\,\pi)^3}\,T^{00}(\vec q^{\,2})\, V^2(\vec q^{\,2}), \label{05}
\end{align}
where 
\begin{align}
V(\vec q^{\,2}) =&\ -\frac{4\,\pi\,Z\,\alpha}{\vec q^{\,2}}\,\rho(\vec q^{\,2}), \label{06}
\end{align}
and $\rho(\vec q^{\,2})$ is the Fourier transform of $\rho(r)$. We will later identify $\rho(-q^2)$ with 
the nuclear charge form factor as a function of the four-momentum square $q^2$ and at the same time
assume that $\rho$ has a proper analytical behavior as a function on the complex plane. 
The two-photon exchange amplitude $T^{\mu\nu}$ is 
\begin{align}
T^{\mu\nu}(p) =&\ \mathrm{Tr}\biggl[\gamma^\mu\,\frac{1}{\not\!p-m}\gamma^\nu\,\frac{\gamma^0+I}{4}\biggr], \label{07}
\end{align}
with $p=t+q$, $t\cdot q=0$, $t=(m,\vec 0)$, and $m$ is the lepton (electron or muon) mass.
Performing the trace
\begin{align}
T^{00}(\vec q^{\,2}) =&\  -\frac{2\,m}{\vec q^{\,2}}\,, \label{08}
\end{align}
$E^{(5,0)}_\mathrm{fns}$ becomes
\begin{align}
E^{(5,0)}_\mathrm{fns} =&\ 
-2\,m\,(4\,\pi\,Z\,\alpha)^2\,\phi^2(0)\,\int \frac{d^3q}{(2\,\pi)^3}\,\frac{\rho^2(\vec q^{\,2})}{\vec q^{\;6}}\,. \label{09}
\end{align}
$E^{(5,0)}_\mathrm{fns}$ requires the subtraction of infrared divergent terms,
which correspond to the point nucleus and the leading $E^{(4)}_\mathrm{fns}$ term; thus
\begin{align}
E^{(5,0)}_\mathrm{fns} =&\ 
-m\,(Z\,\alpha)^2\,\phi^2(0)\!\!\int\! d q\,\frac{16}{q^4}\,\big[\rho^2(q^2)-1-2\,\rho'(0)\,q^2\big]. \label{10}
\end{align}
The $q$ integral can be expressed as 
\begin{align}
\int d q\,\frac{16}{q^4}\,\big[\rho^2(q^2)-1-2\,\rho'(0)\,q^2\big] = r^3_F\,\frac{\pi}{3}, \label{11}
\end{align}
where $r_F$ is defined by \cite{friar:79}
\begin{align}
r^3_F = \int d^3r_1\,d^3r_2\,\rho(\vec r_1)\,\rho(\vec r_2)\,|\vec r_1-\vec r_2|^3, \label{12}
\end{align}
thus
\begin{align}
E^{(5,0)}_\mathrm{fns} =&\ -\frac{\pi}{3}(Z\,\alpha)^2\,\phi^2(0)\,m\,r^3_F, \label{13}
\end{align}
in agreement with Ref. \cite{friar:79}.
This formula is exact only for the nucleus, which is described by the elastic charge form factor $\rho$.
However, the elastic approximation is not always valid, in particular for large momenta $q$. In fact,
values of the electron momenta in Eq.~(\ref{11}) are of the order of the inverse of the nuclear size.
For such high momenta, the electron sees individual nucleons instead of the nucleus as a whole.
It is a small effect for {\em electronic} atoms, because this whole correction is small, namely $O(Z\,\alpha\,m\,r_C)$
in comparison to the leading finite size  $E^{(4)}_\mathrm{fns}$. However, this is not the case of muonic atoms,
where $m\,r_C$ is of order 1 and we find significant inelastic effects \cite{pachucki:24a}.
In general, when large exchange momenta are involved, one may expect a significant inelastic contribution.
The same effect appears at higher-order corrections; a part of fns comes from the low momenta exchange
where the elastic approximation works as for $E^{(4)}_\mathrm{fns}$ and a part from high momenta
where elastic approximation does not work. For example, the so-called three-photon exchange correction
$E^{(6)}_{\rm nucl}$ comes from many scales of exchanged momenta,
and presently only the elastic part  $E^{(6)}_\mathrm{fns}$ is known.
 In this work we assume the elastic approximation, which is usually assumed as a good starting point,
and only in Sec. \ref{VIII} we consider radiative corrections 
to the electric dipole nuclear polarizability effect in {\em electronic} atoms.

\section{$(Z\,\alpha)^5$ Recoil  \MakeLowercase{FNS}}
In contrast to nonrecoil correction from the previous section, the radiative recoil fns corrections
have not yet been extensively studied in the literature. 
Let us first consider the nonradiative recoil fns correction $E^{(5,1)}_\mathrm{fns}$ and follow the derivation from Ref. \cite{pachucki:23}.
According to HPQED the leading  recoil correction \cite{shabaev:85,shabaev:88,pachucki:95a,pachucki:24c} 
in hydrogenic systems, which is exact in $Z\,\alpha$, is of the form
\begin{align}
E_\mathrm{rec} =&\
\frac{i}{M} \int_s \frac{d\omega}{2\,\pi}\,  \langle \phi|D_T^j(\omega) \,G(E_D + \omega)\,D_T^j(\omega)|\phi\rangle \,, \label{14}
\end{align}
where $G(E) = \nicefrac{1}{(E-H_D)}$ and
\begin{align}
D_T^j(\omega,\vec r) =&\ -4\pi Z\alpha \, \alpha^i \, G_{T}^{ij}(\omega,\vec{r})\,, \label{15}
\\
G_T^{ij}(\omega, \vec q) =&\ \frac{\rho(-q^2)}{q^2}\,\biggl(\delta^{ij}-\frac{q^i\,q^j}{\omega^2}\biggr)\,, \label{16}
\end{align}
and where $\omega = q^0$, $q^2 = \omega^2-\vec q^{\,2}$, and
the subscript $s$ in the integration denotes a symmetric  integration around the pole at $\omega=0$ along the Feynman
or Wick rotated contour. Because the terms with $1/\omega$ singularity can be separated from terms involving  branch cuts
starting at $\omega=0$, this symmetric integration can safely be implemented.

The $(Z\,\alpha)^5$ fns recoil beyond the reduced mass scaling of  $\phi^2(0)$ in Eq. (\ref{13}) is
given by the two-photon exchange amplitude
\begin{align}
 E^{(5,1)}_{\rm fns} = &\ 
 \frac{i}{M}\,\phi^2(0)\,(4\,\pi\,Z\,\alpha)^2\! \int_s \frac{d^4q}{(2\,\pi)^4} 
 \nonumber \\ &\times
 G_{T}^{ij}(\omega,\vec{q})\,G_{T}^{ik}(\omega,\vec{q})\,T^{jk}, \label{17}
\end{align}
where $T^{ij}$ is defined in Eq. (\ref{07}).
After contracting Cartesian indices, $ E^{(5,1)}_{\rm fns}$ becomes
  \begin{align}
 E^{(5,1)}_{\rm fns} = &\ 
 \frac{i}{M}\,\phi^2(0)\,(4\,\pi\,Z\,\alpha)^2\! \int_s \frac{d^4q}{(2\,\pi)^4}
  \nonumber \\ &\times
\frac{\omega}{q^2+2\,m\,\omega} \,\biggl(\frac{1}{\omega^4} + \frac{2}{q^4}\biggr)\,\rho^2(-q^2). \label{18}
\end{align}
We perform the Wick rotation of the integration contour $\omega \rightarrow i\,\omega$,
integrate over a three-dimensional sphere (see Appendix A), and subtract the point nucleus and the leading fns,  to obtain
\begin{align}
 E^{(5,1)}_{\rm fns} = &\ 
 -(\Za)^2\,\frac{\phi^2(0)}{m\,M} \int_0^\infty \frac{dq}{q} 
\nonumber \\ &\times
\Big\{ h\Big(\frac{2\,m}{q}\Big)\,\big[1 - \rho^2(q^2)\big]  + \frac{32\,m^3}{q}\,\rho'(0)\Big\}\,, \label{19}
\end{align}
where
\begin{align}
h(a) =&\ 2\,a^2\,\Big[\sqrt{1 + a^2} - \big(1 + \sqrt{1 + a^2}\big)^{-2}\Big]\,. \label{20}
\end{align}
Using the dipole parametrization of the nuclear charge form factor
\begin{align}
\rho(q^2) =&\ \frac{\Lambda^4}{(\Lambda^2+q^2)^2},\;\rho'(0) = -\frac{2}{\Lambda^2},\; r_C^2 = \frac{12}{\Lambda^2}, \label{21}
\end{align}
one obtains \cite{pachucki:25a}
\begin{align}
  E^{(5,1)}_\mathrm{fns} = &\ 
    -\frac{\phi^2(0)}{m\,M}\,(Z\,\alpha)^2
 \biggl[
\frac{3\,(1 - y^2)\,(1 + 4\,y^2 - 35\,y^4)}{8\,y^4}
\nonumber \\ &\
- \frac{3 + 8\,y^2 + 40\,y^4 - 140\,y^6 + 105\,y^8}{16\,y^5}\,\ln\frac{1 + y}{1 - y} 
\nonumber \\ &\
- \ln\frac{1 - y^2}{4}
 \biggr] \label{22}
\,,
\end{align}
where $y = \sqrt{1-4\,m^2/\Lambda^2}$. The expansion of the above formula in small $m/\Lambda$,
which is appropriate for {\em electronic} atoms,  is
\begin{align}
  E^{(5,1)}_\mathrm{fns} = &\
  \frac{\phi^2(0)}{M\,m}\,(Z\,\alpha)^2\biggl[\bigg(\! \frac{43}{12} - \ln 12 + 2\,\ln m\,r_C\! \bigg) (m\,r_C)^2
\nonumber \\ &\ 
- \frac{25}{36}\biggl(\frac{3}{20} - \ln 12 + 2\,\ln m\,r_C\!\biggr)\,(m\,r_C)^4
\nonumber \\ &\ 
+ O\big(m\,r_C\big)^6\biggr]\,. \label{23}
\end{align}
For muonic atoms, $E^{(5,1)}_\mathrm{fns}$ is a significant correction, but $m\,r_C$ 
can be close to 1; therefore, the formulas Eq. (\ref{19}) or (\ref{22})  should be used, instead of the  $m\,r_C$ expansion.
Moreover, one can expect significant inelastic contributions. 
In fact, for $\mu$H one uses dispersion relations to express the exact two-photon exchange amplitude 
in terms of the measured inelastic structure functions \cite{carlson:11}.

\section{$\alpha\,(Z\,\alpha)^5$ radiative nonrecoil  \MakeLowercase{FNS}}
The $\alpha\,(Z\,\alpha)^5$ radiative correction $E^{(6,0)}_\mathrm{radfns}$ to the finite nuclear size effect
is obtained by a replacement in Eq. (\ref{05}) of $T^{00}$ with radiatively corrected $T^{00}$
\begin{align}
E^{(6,0)}_\mathrm{radfns} =&\ 
\phi^2(0)\!\int \frac{d^3p}{(2\,\pi)^3}\, \big[T^{00}_\mathrm{se}(\vec p^{\,2}) +T^{00}_\mathrm{vp}(\vec p^{\,2})\big]\,V^2(\vec p^{\,2}),
\label{24}
\end{align}
where $T^{00}_\mathrm{se}(\vec p^{\,2})$ is the electron self-energy correction to $T^{00}(\vec p^{\,2})$ presented in Eq. (\ref{C05}) of Appendix C, 
and $T^{00}_\mathrm{vp}(\vec p^{\,2})$ is the  vacuum-polarization correction to $T^{00}(\vec p^{\,2})$
\begin{align}
T^{00}_\mathrm{vp}( p^{\,2}) =&\  \frac{4}{p^2}\, \bar\omega\bigg(\frac{p^{\,2}}{m^2}\bigg), \label{25}
\end{align}
where
\begin{align}
\bar\omega\bigg(\frac{p^2}{m^2}\bigg) =&\  -\frac{\alpha}{3\,\pi}\,\biggl\{\frac{1}{3} + 2\biggl(1-\frac{2\,m^2}{p^2}\biggr)
\nonumber \\ \times&\
\biggl[\sqrt{1+\frac{4\,m^2}{p^2}}\,\mathrm{arccoth}\sqrt{1+\frac{4\,m^2}{p^2}}-1\biggr]\biggr\}. \label{26}
\end{align}
Let us rescale all momenta by the lepton mass $m$ in Eq. (\ref{24}), subtract the point nucleus,
and introduce a function $f$
\begin{align}
E^{(6,0)}_\mathrm{radfns} =&\ \frac{\alpha}{\pi}\,\frac{\phi^2(0)}{m^2}\,\int\frac{d^3p}{(2\,\pi)^3}\,\frac{(4\,\pi\,Z\,\alpha)^2}{p^4}
\nonumber \\ &\times
f(p^2)\,\left[\rho\big(m^2\,p^2\big)^2-1\right], \label{27}
\end{align}
where
\begin{align}
\frac{\alpha}{\pi}\,f(p^2) = T^{00}_\mathrm{se}(p^{2}) +T^{00}_\mathrm{vp}(p^{2})\,. \label{28}
\end{align}
The function $f(p^2)$ can be obtained analytically, but following Ref. \cite{pachucki:93a} it is more convenient to use its imaginary part 
on the branch cut 
\begin{align}
f^A(q^2) =&\ \frac{ f(-q^2 +i\,\epsilon) - f(-q^2 -i\,\epsilon)}{2\,\pi\,i}\,, \label{29}
\end{align}
so that
\begin{align}
f(p^2) =&\ -\int_0^\infty d(q^2)\,\frac{f^A(q^2)}{q^2+p^2}\,, \label{30}
\end{align}
where
\begin{align}
f^A(q^2) =&\ 
-\frac{3}{4} - \frac{4}{q^2} - \frac{1}{4\,(1 + q^2)} - J^A\,\biggl(1 + \frac{4}{q^2} - \frac{q^2}{4}\biggr)  
\nonumber \\ &\
+ \frac{\Theta(q-2)}{\sqrt{1 - \frac{4}{q^2}}}\, \biggl(1 +\frac{1}{q^2} - \frac{12}{q^4} \biggr)
\nonumber \\ &\
+ \frac{4}{3}\, \Theta(q-2) \sqrt{1-\frac{4}{q^2}}\,\frac{1}{q^2}\, \bigg(1+\frac{2}{q^2}\bigg)\,, \label{31}
 \end{align}
in which the last term comes from the vacuum polarization, and
\begin{align}
J^A = -\frac{1}{q}\,\bigg[\arctan(q)-\Theta(q-2)\, \arccos\bigg(\frac{2}{q}\bigg)\bigg]\,. \label{32}
\end{align}
For muonic atoms, the integral in Eq. (\ref{27}) should be performed numerically without expansion 
of $\rho(p^2)$ in small momenta; this was done in Ref. \cite{pachucki:24a} for 
$\mu$H, $\mu$D, $\mu^3$He, and $\mu^4$He.
For {\em electronic} atoms the finite nuclear size is much smaller than the electron Compton wavelength, 
so one can perform small $m\,r_C$ expansion
\begin{align}
E^{(6,0)}_\mathrm{radfns} =&
 \alpha\,(Z\,\alpha)^2\,\frac{\phi^2(0)}{m^2}\,
\biggl[
\frac{2\,\pi}{3}\,(m\,r_C)^2\,(4\,\ln 2 -5)
\nonumber \\ &\
+ \frac{(m\,r_F)^3}{9}\,\biggl(\frac{9019}{1260} - \ln 12 + 2\ln m\,r_C \biggr)
\nonumber \\ &\ 
+ \frac{2\,\pi}{3}\,\bigg[ \frac{(m\,r_C)^4}{3} + \frac{(m\,r_{CC})^4}{5}\bigg]\,\biggl( 4\,\ln 2 - \frac{115}{32} \biggr) 
\nonumber \\ &\ + \ldots \biggr]. \label{34}
\end{align}
The first term $4\,\ln 2-5 = -6.997\,619/\pi$ is in agreement with the result from Eides \cite{eides:01}, 
and we correct here our original result in Ref. \cite{pachucki:93a}, which was $-6.744\,02/\pi$.
The second term in the above equation is obtained for the dipole parametrization of the nuclear charge form factor.

There are also higher-order in $Z\,\alpha$ radiative corrections, studied in Ref. \cite{sgk:18}
using analytic methods and in Ref. \cite{yerokhin:11} using numerical methods. Here we concentrate on 
nuclear recoil effects, which is the topic of the next section.
 
\section{$\alpha\,(Z\,\alpha)^5$ radiative RECOIL \MakeLowercase{FNS}}
We will consider the radiative recoil effects separately for the vacuum polarization and the lepton self-energy 
\begin{align}
 E^{(6,1)}_{\rm radfns} =&\   E^{(6,1)}_{\rm vpfns} +  E^{(6,1)}_{\rm sefns}. \label{35}
 \end{align}
$E^{(6,1)}_{\rm vpfns}$  is obtained by multiplying the integrand of Eq. (\ref{18}) by $-2\,\bar\omega$
and subtracting a point nucleus
\begin{align}
 E^{(6,1)}_{\rm vpfns} =&\ 
(\Za)^2\,\frac{\phi^2(0)}{m\,M} \int_0^\infty \frac{dq}{q}
\nonumber \\ &\times
2\,\bar\omega\Big(\frac{q^2}{m^2}\Big)
h\Big(\frac{2\,m}{q}\Big)\,\big[1 - \rho^2(q^2)\big] \label{36}
\end{align}
because the muon Compton wave length is much smaller than the Bohr radius, and thus no contribution
of lower order is present.
For muonic atoms $E^{(6,1)}_{\rm vpfns}$ will be combined with $E^{(6,1)}_{\rm sefns}$, while
for {\em electronic} atoms it can be further simplified to
\begin{align}
 E^{(6,1)}_{\rm vpfns} =&\ 
-(\Za)^2\,\phi^2(0)\,\frac{m\,}{M}\,\frac{\alpha}{\pi}\,\frac{r_C^2}{12}
\nonumber \\ &\times
\bigg(\frac{547}{9} + \frac{4\,\pi^2}{3} + 22\,\ln\frac{m^2\,r_C^2}{12} + 4\,\ln^2\frac{m^2\,r_C^2}{12} \bigg). \label{37}
\end{align}
We note that the $\ln^2$ term cancels out with the corresponding self-energy correction.

$E^{(6,1)}_{\rm sefns}$ is obtained from Eq. (\ref{17}) by replacement of $T^{jk}$ by 
the self-energy corrected $T^{jk}_\mathrm{se}$
\begin{align}
 E^{(6,1)}_{\rm sefns} = &\ 
 \frac{i}{M}\,\phi^2(0)\,(4\,\pi\,Z\,\alpha)^2
 \nonumber \\ &\times
 \int_s \frac{d^4q}{(2\,\pi)^4} 
 \,G_{T}^{ij}(\omega,\vec{q})\,G_{T}^{ik}(\omega,\vec{q})\,T_\mathrm{se}^{jk}(q)\,. \label{38}
\end{align}
Because the symmetrized tensor 
\begin{equation}
T_\mathrm{sym}^{\mu\nu}(q) = T_\mathrm{se}^{\mu\nu}(q)+T_\mathrm{se}^{\mu\nu}(-q) \label{39}
\end{equation}
 fulfills the continuity relation $ q_\mu\,T_\mathrm{sym}^{\mu\nu}(q) = 0$, the contraction of indices leads to
\begin{align}
T_\mathrm{sym}^{jk}\,\biggl(\delta^{ij}-\frac{q^i\,q^j}{\omega^2}\biggr)\, \biggl(\delta^{ik}-\frac{q^i\,q^k}{\omega^2}\biggr)
=&\
-T^\mu_{\mathrm{sym},\,\mu} - \frac{q^2}{\omega^2}\,T_\mathrm{sym}^{00}\,, \label{40}
\end{align}
and thus with $\omega = q^0$
\begin{align}
 E^{(6,1)}_{\rm sefns} =&\
 \frac{1}{M}\,\phi^2(0)\,(4\,\pi\,Z\,\alpha)^2\! \int_s \frac{d^4q}{(2\,\pi)^4\,i}
 \nonumber \\ &\hspace{-5ex}\times
 \bigg(\frac{T^\mu_{\mathrm{se},\,\mu}(-q^2,q^0)}{q^2} 
 + \frac{T_\mathrm{se}^{00}(-q^2,q^0)}{q_0^2}\bigg)\,\frac{\rho^2(-q^2)-1}{q^2}\,, \label{41}
 \end{align}
where $T_\mathrm{se}^{00}$ and $T^\mu_{\mathrm{se},\,\mu}$ are presented in Appendix C.
For the numerical calculation it is convenient to perform Wick rotation, and 
using the Euclidean metrics
 \begin{align}
 E^{(6,1)}_{\rm sefns} \stackrel{E}{=}&\
   \frac{1}{M}\,\phi^2(0)\,(4\,\pi\,Z\,\alpha)^2\! \int \frac{d^4q}{(2\,\pi)^4}
\nonumber \\ &\hspace{-5ex}\times   
  \bigg(\frac{T^\mu_{\mathrm{se},\,\mu}(q^2,i\,q^0) }{q^2} 
 + \frac{T_\mathrm{se}^{00}(q^2,i\,q^0)}{q_0^2}\bigg)\,\frac{\rho^2(q^2)-1}{q^2}. \label{42}
 \end{align}
 The above integrals are calculated by introducing a dimensionless $T$
 \begin{align}
 E^{(6,1)}_{\rm sefns} =&\ \frac{\alpha}{\pi}\,(Z\,\alpha)^2\,\frac{\phi^2(0)}{M\,m}\,\frac{T}{2},  \label{43}
 \end{align}
 which is defined as
  \begin{align}
T =&\ \int_0^\infty \frac{dq}{q}\,\big[\rho^2(q^2)-1\big]\, T(q),  \label{44}
\end{align}
where
\begin{align}
 T(q) =&\  \frac{2}{\pi}\int_0^\pi d\phi\,\sin^2\phi\, \bigg(T^\mu_{\mathrm{se},\,\mu}(q^2,i\,q\,\cos\phi) 
\nonumber \\ &\ 
 + \frac{T_\mathrm{se}^{00}(q^2,i\,q\,\cos\phi)}{\cos^2\phi}\bigg)\,\frac{4\,\pi}{\alpha}.  \label{45}
 \end{align}
The numerical integration of the above double integral is performed as follows.
We integrate over $\phi$ by parts, using Eq. (\ref{B02}) from Appendix B, to replace $J$
with its derivative $\partial J/\partial\phi$, so only logs are present in the integrand.
Next, the terms with $1/\cos^2(\phi)$ are separated out and integrated over $\phi$ analytically.
The numerical integration of the remaining terms over $\phi$ does not have any singular points, 
but requires adaptive quadrature and is performed with the {\sl Mathematica} integration routine.
 The second integral over $q$ is smooth, but the integrand contains $\ln q$ at 
 both integration boundaries at  $q=0$ and $q=\infty$ and a spurious singularity at $q=1$.
 Therefore, this integral is split into two parts 
 \begin{align}
T =&\  \int_0^1 \frac{dq}{q}\,\big[\rho^2(q^2)-1\big]\, T(q) 
\nonumber \\ &\
+ \int_0^1 \frac{dx}{x}\,\big[\rho^2(1/x^2)-1\big]\, T(1/x)  \label{46}
\end{align} 
and is performed using the extended Gaussian quadrature \cite{pachucki:14},
which is adapted to polynomials and logarithm times polynomials at the same time.
Its numerical convergence is very fast, and with only 20 integration points we got about 16 significant digits. 
For {\em electronic} atoms one performs a small $m\,r_C$ expansion and obtains
 \begin{align}
 E^{(6,1)}_{\rm sefns} =&\
 \frac{m}{M}\,\frac{\alpha}{\pi}\,(Z\,\alpha)^2\,\phi^2(0)\,   
  \frac{r_C^2}{12}\,\biggl[ 261.253\,141
  \nonumber \\ &\
+ \frac{7}{3}\,\ln\biggl(\frac{m^2\,r_C^2}{12}\biggr) + 4\,\ln^2\biggl(\frac{m^2\,r_C^2}{12}\biggr)\biggr]\,.  \label{47}
\end{align}
As previously noted, the $\ln^2$ term cancels out with the corresponding vp contribution.
 
 \section{EVP   \MakeLowercase{FNS}}
There is an additional fns correction for muonic atoms
that comes from the electron vacuum polarization (evp) potential, 
\begin{align}
V_\mathrm{evp} =&\ 4\,\pi\,Z\,\alpha\,\int\frac{d^3q}{(2\,\pi)^3}\,\frac{\,e^{i\,\vec q\cdot\vec r}}{\vec q^{\,2}}
\,\bar\omega\bigg(\frac{\vec q^{\,2}}{m_e^2}\bigg). \label{48}
\end{align}
Let us now introduce $\eta_\mathrm{evp}$ defined by
\begin{align}
 \frac{\alpha}{\pi}\,\eta_\mathrm{evp} =&\ \frac{\phi^2_\mathrm{vp}(0)}{\phi^2(0)} = 
 \frac{2}{\langle \delta^3(r)\rangle}\, \bigg\langle\!\delta^3(r)\,\frac{1}{(E-H)'}\,V_\mathrm{evp}\!\bigg\rangle. \label{49}
 \end{align}
 It can be calculated semianalytically for arbitrary states of light muonic atoms \cite{sgk:18}.
 A large part of evp-fns correction is obtained by $\eta_\mathrm{evp}$ rescaling of the wave function at the origin
 \begin{align}
 E^{\eta}_\mathrm{evpfns}=&\ \frac{\alpha}{\pi}\,\eta_\mathrm{evp}\,\big(E^{(4)}_\mathrm{fns} + E^{(5,0)}_\mathrm{fns} + E^{(5,1)}_\mathrm{fns} + \ldots\big), \label{50}
 \end{align}
 Beyond this rescaling  the evp-fns correction  in the leading order of $Z\,\alpha$ has the form \cite{pachucki:24a}
\begin{align}
E^{(5)}_\mathrm{evpfns} =&\ \frac{r_C^2}{6}\,\big\langle\nabla^2V_\mathrm{evp}\big\rangle. \label{51}
\end{align}
The higher-order evp correction in the infinite nuclear mass limit, $E^{(6,0)}_\mathrm{evpfns}$, 
is given by the two-photon exchange amplitude
\begin{align}
E^{(6,0)}_\mathrm{evpfns} =&\
(Z\,\alpha)^2\,\phi^2(0)\,m\int d q\,2\,\bar\omega\bigg(\frac{q^2}{m_e^2}\bigg)
\nonumber \\ &\times
\frac{16}{q^4}\,\big[\rho(q^2)^2-1-2\,\rho'(0)\,q^2\big]\,, \label{52}
\end{align}
with subtracted $\rho'(0)$, which corresponds to the lower-order term in Eq. (\ref{51}).  
Using a dipole parametrization of the nuclear form factor
and a large $q^2$ asymptotics of $\bar\omega$
\begin{align}
\bar\omega\bigg(\frac{q^2}{m_e^2}\bigg) \approx&\ \frac{\alpha}{3\,\pi}\,\biggl(\frac{5}{3} + \ln\frac{m_e^2}{q^2} \biggr), \label{53}
\end{align}
one obtains
\begin{align}
E^{(6,0)}_\mathrm{evpfns} =&\ 
(Z\,\alpha)^2\,\phi^2(0)\,m\int d q\,
\frac{2\,\alpha}{3\,\pi}\,\biggl(\frac{5}{3} + \ln\frac{m_e^2}{q^2} \biggr)
\nonumber \\ &\times
\frac{16}{q^4}\,\big[\rho^2(q^2)-1-2\,\rho'(0)\,q^2\big]
\nonumber \\ =&\
-\frac{2\,\alpha}{3\,\pi}\,\bigg(2\,\ln(m_e\,r_C)  +\frac{811}{315} - \ln 12 \bigg)\,E^{(5,0)}_\mathrm{fns},  \label{54}
\end{align}
in agreement with Eq. (26) of Ref. \cite{sgk:18}, and we note that the constant term
is obtained for the dipole parametrization of the nuclear charge form factor. 

Let us pass now to calculation of the electron vacuum polarization with recoil and with finite nuclear size correction in muonic atoms.
The evp-rec-fns correction beyond the reduced mass scaling of $\phi^2(0)$, using Eq. (\ref{19}), is
\begin{align}
 E^{(6,1)}_{\rm evpfns} =&\ 
(\Za)^2\,\frac{\phi^2(0)}{m\,M} \int_0^\infty \frac{dq}{q}\,2\,\bar\omega(q^2/m_e^2)
\nonumber \\ &\times
\Big\{h(a)\,\big[1 - \rho^2(q^2)\big]  + 16\,m^2\,a\,\rho'(0)\Big\}\,, \label{55}
\end{align}
where $m$ is the muon mass,  $a=2\,m/q$, and $\bar\omega$ is defined in Eq.~(\ref{26}). 
Using the large $k$ asymptotics of $\bar\omega(q^2/m_e^2)$, $E^{(6,1)}_{\rm evpfns}$ becomes
\begin{align}
 E^{(6,1)}_{\rm evpfns} =&\ 
(\Za)^2\,\frac{\phi^2(0)}{m\,M} \int_0^\infty \frac{dq}{q}\,\frac{2\,\alpha}{3\,\pi}\,\biggl(\frac{5}{3} +\ln\frac{m_e^2}{q^2}\biggr)
\nonumber \\ &\times
\Big\{h(a)\,\big[1 - \rho^2(q^2)\big]  + 16\,m^2\,a\,\rho'(0)\Big\}\,. \label{56}
\end{align}
This integral will be performed numerically together with all other corrections from previous sections.

\section{Radiative correction to the nuclear polarizability effect \label{VIII}}

Perhaps the most interesting are radiative corrections to the nuclear polarizability effect in atomic systems,
which have not yet been investigated in the literature. Calculations proceed similarly
to the previously considered radiative recoil fns correction, and let us first recall, at the beginning,
the derivation of the leading electric dipole nuclear polarizability correction in {\em electronic} atoms. 
The interaction of a nucleus with the electric field is given by
\begin{align}
\delta H = Z\,e\,A^0 -\vec d\cdot\vec E,
\end{align}
where $\vec d$ is the electric dipole operator. The electric field from the electron can excite
the nucleus, which effectively changes atomic energy levels. The shift, 
at the leading order, is given by the two-photon exchange in the scattering approximation,
which in the temporal gauge takes the form
\begin{align}
E_{\rm pol} =&\
i\,e^2\,\phi^2(0)\,\int\frac{d^4q}{(2\,\pi)^4}\,\omega^2\,
\frac{\Bigl(\delta^{ik}-\frac{q^i\,q^k}{\omega^2}\Bigr)}{q^2}\,
\frac{\Bigl(\delta^{jl}-\frac{q^j\,q^l}{\omega^2}\Bigr)}{q^2}
\nonumber \\ & \hspace*{-5ex} \times
\big[T^{ij}(q) + T^{ji}(-q)\big]\,
\biggl\langle d^{\,k}\,\frac{1}{E_N-H_N-\omega}\,d^{\,l}\biggr\rangle, \label{pol58}
\end{align}
where $\omega = q^0$ and $T^{ij}$ is defined in Eq. (\ref{07}).
Let us perform the $q$ integration in the way, that is  suited for later calculations of radiative corrections.
Therefore, we assume the Wick rotated contour $\omega \rightarrow i\,\omega$, use the Euclidean metric 
\begin{widetext}
\begin{align}
E_{\rm pol} =&\
e^2\,\phi^2(0)\,\int\frac{d^4q}{(2\,\pi)^4}\,\frac{4\,m}{3}\,
\frac{q^4 + 2\,\omega^4}{q^4\,(q^2 -2\,i\,m\,\omega)\,(q^2 + 2\,i\,m\,\omega)}\,
\biggl\langle \vec d\,\frac{1}{E_N-H_N-i\,\omega}\,\vec d\biggr\rangle\,,
\end{align}
perform the angular average
\begin{align}
E_{\rm pol} =&\
-e^2\,\phi^2(0)\,\int\frac{d^4q}{(2\,\pi)^4}\,\frac{4\,m}{3}\,\int\frac{d\,\Omega_q}{2\,\pi^2}\frac{q^4 + 2\,q_0^4}{q^4\,(q^4 + 4\,m^2\,q_0^2)}\,
\biggl\langle \vec d\,\frac{1}{H_N-E_N+i\,q^0}\,\vec d\biggr\rangle\,, \label{pol60}
\end{align}
 and expand in the small ratio of the electron mass to the nuclear excitation energy
\begin{align}
E_{\rm pol} =&\ -m\,\alpha^2\,\phi^2(0)\,\frac{2}{3}\,
 \biggl\langle \frac{\vec d}{e}\,\frac{1}{H_N-E_N}\,\biggl[\frac{19}{6} + 5\,\ln\frac{2\,(H_N-E_N)}{m} \biggr]\,\frac{\vec d}{e}\biggr\rangle\,.
\end{align}
The corresponding contribution to the $2S$--$1S$ transition in ordinary deuterium is significant and amounts to 
$19.26\,(6)$~kHz \cite{friar:97:a}.
Calculation of the electron vacuum polarization correction proceeds in a similar way. Using Eq. (\ref{pol60})
the evp correction is
\begin{align}
E_{\rm vppol} =&\
-e^2\,\phi^2(0)\,\int\frac{d^4q}{(2\,\pi)^4}\,\frac{4\,m}{3}\,\int\frac{d\,\Omega_q}{2\,\pi^2}\frac{q^4 + 2\,q_0^4}{q^4\,(q^4 + 4\,m^2\,q_0^2)}\,
\biggl\langle \vec d\,\frac{1}{H_N-E_N+i\,q^0}\,\vec d\biggr\rangle\,
(-2)\,\bar\omega\biggl(\frac{q^2}{m^2}\biggr).
\end{align}
The expansion in the small ratio of the electron mass to the nuclear excitation energy leads to
\begin{align}
E_{\rm vppol} =&\
m\,\alpha^2\,\phi^2(0)\,\frac{\alpha}{\pi}\,
 \biggl\langle \frac{\vec d}{e}\,\frac{1}{H_N-E_N}\,
 \biggl[-5.900\,723\,922 - \frac{62}{27}\,\ln\frac{2\,(H_N-E_N)}{m} 
 - \frac{20}{9}\,\ln^2\frac{2\,(H_N-E_N)}{m}
 \biggr]\,\frac{\vec d}{e}\biggr\rangle. 
\end{align}

The electron self-energy correction is obtained by replacement of $T^{ij}$ with $T^{ij}_\mathrm{se}$ in Eq. (\ref{pol58}), so
\begin{align}
E_{\rm sepol} =&\
i\,e^2\,\phi^2(0)\,\,\int\frac{d^4q}{(2\,\pi)^4}\,\frac{\omega^2}{q^4}\,
\Bigl(\delta^{ik}-\frac{q^i\,q^k}{\omega^2}\Bigr)\,
\Bigl(\delta^{jk}-\frac{q^j\,q^k}{\omega^2}\Bigr)\,\big[T^{ij}_\mathrm{se}(q) + T^{ij}_\mathrm{se}(-q)\big]
\biggl\langle \vec d\,\frac{1}{E_N-H_N-\omega}\,\vec d\biggr\rangle\,\frac{1}{3}.
\end{align}
Using the continuity condition  and Eq. (\ref{40}),
\begin{align}
E_{\rm sepol} =&\
-i\,e^2\,\phi^2(0)\,\int\frac{d^4q}{(2\,\pi)^4}\,\frac{1}{q^4}\,
\bigl(\omega^2\,T^\mu_{\mathrm{se}, \mu}+ q^2\,T^{00}_\mathrm{se}\bigr)
\biggl[\biggl\langle\!\vec d\,\frac{1}{E_N-H_N-\omega}\,\vec d\biggr\rangle
+ \biggl[\biggl\langle\! \vec d\,\frac{1}{E_N-H_N+\omega}\,\vec d\biggr\rangle\biggr]\,\frac{1}{3},
\end{align}
where $T^\mu_{\mathrm{se}}$ and $T^{00}_\mathrm{se}$ are presented in Appendix C.
Further calculations proceed similarly to that of radiative recoil fns. 
Using Eq. (\ref{B02}) we integrate by parts to eliminate $J$
and perform expansion in  the small ratio of the electron mass to the nuclear excitation energy
\begin{align}
E_{\rm sepol} =&\
m\,\alpha^2\,\phi^2(0)\,\frac{\alpha}{\pi}\,
 \biggl\langle \frac{\vec d}{e}\,\frac{1}{H_N-E_N}\,
 \biggl[-0.345\,584\,311 + \frac{9}{2}\,\ln\frac{2\,(H_N-E_N)}{m} 
 + \frac{8}{9}\,\ln^2\frac{2\,(H_N-E_N)}{m}
 \biggr]\,\frac{\vec d}{e}\biggr\rangle.
\end{align}
\end{widetext}
One notes that radiative corrections to the nuclear polarizability effect are decreased by a factor $\alpha/\pi$ and are slightly enhanced
by the presence of an additional logarithm in the nuclear excitation energy.

\section{Results and Summary}
\begin{table}[t]
\renewcommand{\arraystretch}{0.92}
\caption{$\alpha\,(Z\,\alpha)^5\,m^2/M$ corrections for the $2P$-$2S$ transition in (two-body) muonic atoms in meV units,
$\delta E_\mathrm{exp}$ is the current experimental uncertainty, 
and $\eta_\mathrm{evp}$ is taken from Table 13 of Ref. \cite{sgk:21}}
\label{table_mu}
\begin{ruledtabular}
\begin{tabular}{lw{1,8}w{2.8}w{2.8}w{2.8}}
&\multicolumn{1}{c}{$\mu$H} &   \multicolumn{1}{c}{$\mu$D}
&  \multicolumn{1}{c}{$\mu^3$He} &  \multicolumn{1}{c}{$\mu^4$He}\\[0.5ex] \hline \\[-1.5ex]
$r_C$[fm] & 0.840\,60(39)       &   2.127\,58(78)&	1.970\,07(94) & 1.678\,6(12)	\\
$m_\mu\,r_C$ & 0.450\,1	       &    1.139\,2   &         1.054\,9  & 0.898\,8   \\	
$m_\mu/m_N$ & 0.112\,610 & 0.056\,333 & 0.037\,622 & 0.028\,347 \\
$\eta_\mathrm{evp}(2S)$ & 1.404\,1 & 1.452\,3 & 2.181\,8 & 2.192\,0\\[1.5ex]
$\delta E_\mathrm{exp}$ & 0.002\,3 & 0.003\,4 & 0.048 & 0.058 \\[0.5ex]
$E^{(6,1)}_\mathrm{vpfns}$ & 0.000\,01 & 0.000\,02 & 0.000\,41 &  0.000\,27 \\[0.5ex]
$E^{(6,1)}_\mathrm{sefns}$ & -0.000\,08 & -0.000\,23 & -0.004\,49 & -0.002\,67 \\[0.5ex]
$E^{(6,1)}_\mathrm{evpfns}$ &-0.000\,01 & -0.000\,32 & -0.005\,72 & -0.002\,61\\[0.5ex]
$\frac{\alpha}{\pi}\eta_\mathrm{evp} E^{(5,1)}_\mathrm{fns}$ & -0.000\,01 &  -0.000\,06 & -0.001\,94 & -0.001\,06 
\end{tabular}
\end{ruledtabular}
\end{table}
Let us now summarize the obtained results.
The total radiative nonrecoil FNS correction at $\alpha\,(Z\,\alpha)^5\,m$ order for muonic atoms is the sum
\begin{align}
E^{(6,0)}_\mathrm{radfns} + E^{(6,0)}_\mathrm{evpfns} + \frac{\alpha}{\pi}\,\eta_\mathrm{evp}\,E^{(5,0)}_\mathrm{fns}, \label{57}
\end{align}
where
\begin{align}
E^{(6,0)}_\mathrm{radfns} =&\ \frac{\alpha}{\pi}\,\frac{\phi^2(0)}{m^2}\,\int\frac{d^3p}{(2\,\pi)^3}\,\frac{(4\,\pi\,Z\,\alpha)^2}{p^4}
\nonumber \\ &\times 
f(p^2)\,\left[\rho^2\big(m^2\,p^2\big)-1\right],  \label{58}\\
E^{(6,0)}_\mathrm{evpfns} =&\ 
-\frac{2\,\alpha}{3\,\pi}\,\bigg(2\,\ln(m_e\,r_C)  +\frac{811}{315} - \ln 12 \bigg)\,E^{(5,0)}_\mathrm{fns},\label{59}
\end{align}
and where $\eta_\mathrm{evp}$ is defined in Eq. (\ref{49}) and $E^{(5,0)}_\mathrm{fns}$ in Eq. (\ref{13}).
All the above corrections are significant in muonic atoms but shall be treated slightly differently.
The first one, $E^{(6,0)}_\mathrm{radfns}$, is dominated by low momenta, so the elastic approximation is appropriate,
and Eq. (\ref{58}) has been used as it stands for $\mu$H, $\mu$D, $\mu^3$He, and $\mu^4$He in Ref. \cite{pachucki:24a}.
The second correction is dominated by large momenta because it is proportional to $r^3_F$;
thus Eq. (\ref{59}) has not been used. Instead we calculated the evp correction to the complete two-photon exchange contribution;
see Ref. \cite{pachucki:24a} for details.

For {\em electronic}  atoms, the expansion of  $E^{(6,0)}_\mathrm{radfns}$  in small  $m\,r_C$ gives
\begin{align}
E^{(6,0)}_\mathrm{radfns} =&\ 
\alpha\,(Z\,\alpha)^2\,\frac{\phi^2(0)}{m^2}\,
\biggl[
\frac{2\,\pi}{3}\,(m\,r_C)^2\,(4\,\ln 2 -5)
\nonumber \\ &\
+ \frac{(m\,r_F)^3}{9}\,\biggl(\frac{9019}{1260} - \ln 12 + 2\ln m\,r_C \biggr)
\nonumber \\ &\ 
+ \frac{2\,\pi}{3}\,\bigg[ \frac{(m\,r_C)^4}{3} + \frac{(m\,r_{CC})^4}{5}\bigg]\,\biggl( 4\,\ln 2 - \frac{115}{32} \biggr)   \label{60}
\biggr].
\end{align}
The first term agrees with the former result of Eides {\em et al.} \cite{eides:01}, and this correction has been included in all
modern determinations of nuclear charge radii from the isotope shift measurements.
 
The next result of this work is the derivation of the total radiative recoil fns correction, which for muonic atoms is the sum
\begin{align}
E^{(6,1)}_\mathrm{radfns} + E^{(6,1)}_\mathrm{evpfns} + \frac{\alpha}{\pi}\,\eta_\mathrm{evp}\,E^{(5,1)}_\mathrm{fns}, \label{61}
\end{align}
where
\begin{align}
E_\mathrm{radfns}^{(6,1)} =&\  E_\mathrm{vpfns}^{(6,1)} + E_\mathrm{sefns}^{(6,1)}
\nonumber \\ =&\  
\frac{\alpha}{\pi}\,(\Za)^2\,\frac{\phi^2(0)}{m\,M}\,\int\frac{dq}{q}
\bigl[{\cal E}_\mathrm{vpfns}^{(6,1)}(q) + {\cal E}_\mathrm{sefns}^{(6,1)}(q)\bigr],  \label{62}
\\
E_\mathrm{evpfns}^{(6,1)} =&\
\frac{\alpha}{\pi}\,(\Za)^2\,\frac{\phi^2(0)}{m\,M}\,\int\frac{dq}{q} {\cal E}_\mathrm{evpfns}^{(6,1)}(q),  \label{63}
\end{align}
and where $E^{(5,1)}_\mathrm{fns}$ is defined in Eq. (\ref{19}),
\begin{align}
{\cal E}_\mathrm{vpfns}^{(6,1)}(q)  =&\ 
\frac{2\,\pi}{\alpha}\,\bar\omega\Big(\frac{q^2}{m^2}\Big)
h\Big(\frac{2\,m}{q}\Big)\,\big[1 - \rho^2(q^2)\big],  \label{64}
\\
{\cal E}^{(6,1)}_{\rm sefns}(q) =&\ -\frac{1}{2}\, T(q) \,\big[1-\rho^2(q^2)\big],  \label{65}
\\
{\cal E}_\mathrm{evpfns}^{(6,1)}(q) =&\ 
\frac{2}{3}\,\biggl(\frac{5}{3} +\ln\frac{m_e^2}{q^2}\biggr)\,\bigg\{
\frac{32\,m^3}{q}\,\rho'(0)
\nonumber \\ &\
+h\Big(\frac{2\,m}{q}\Big)\,\big[1 - \rho^2(q^2)\big]\bigg\}\,.  \label{66}
\end{align}
In the above,  $\rho$ is the nuclear charge form-factor, $h$ is defined in Eq. (\ref{20}), $\bar\omega$ in Eq. (\ref{26}), and $T$ in Eqs. (\ref{49},\ref{50}).
Numerical results for  $\mu$H, $\mu$D, $\mu^3$He, and $\mu^4$He are presented in Table I.
They happened to be well below the current experimental uncertainties.  
Most probably, these corrections will be more important for the hyperfine splitting,
especially for $\mu$H, measurement of which is under preparation \cite{nuber:23, vacchi:23}.

For {\em electronic}  atoms, this correction takes the form
\begin{align}
E_\mathrm{radfns}^{(6,1)} =&\ 
\frac{m}{M}\,\frac{\alpha}{\pi}\,(Z\,\alpha)^2\,\phi^2(0)\,r_C^2
\nonumber \\ &\ \times
\biggl(- \frac{59}{18}\,\ln m\,r_C + 19.682\,143\biggr). \label{67}
\end{align}
One notes that the quadratic logarithmic terms cancel out between self-energy and vacuum-polarization contributions.  
For this reason, the radiative recoil fns correction is quite small, namely below 1Hz for the ground state of the hydrogen atom, 
and thus is negligible in {\em electronic} atoms.

The last result of this work is the radiative correction to the nuclear polarizability effect in {\em electronic} atoms,
\begin{align}
E_{\rm radpol} =&\ E_{\rm vppol}  + E_{\rm sepol} 
\nonumber \\ &\hspace*{-10ex} 
= m\,\alpha^2\,\phi^2(0)\,\frac{\alpha}{\pi}\,
 \biggl\langle\frac{\vec d}{e}\,\frac{1}{H_N-E_N}\,
 \biggl[-6.246\,308
 \nonumber \\ & \hspace*{-10ex}
 + \frac{367}{54}\,\ln\frac{2\,(H_N-E_N)}{m} 
 - \frac{4}{3}\,\ln^2\frac{2\,(H_N-E_N)}{m}
 \biggr]\,\frac{\vec d}{e}\biggr\rangle,
\end{align} 
Let us estimate it for the $2S-1S$ transition in D using the calculated 
in Ref. \cite{pachucki:93b} mean deuteron excitation energy $\langle E\rangle = 7.141$ MeV, so
$\eta = \ln\frac{2\,\langle E\rangle}{m} = 3.330$ and
 \begin{align}
E_{\rm radpol} =&\ E_{\rm pol}\,\frac{\alpha}{\pi}\,
\frac{6.246 - \frac{367}{54}\,\eta + \frac{4}{3}\,\eta^2}{\frac{2}{3}\,\bigl(\frac{19}{6} + 5\,\eta\bigr)}
\nonumber \\
=&\ - E_{\rm pol}\,\frac{\alpha}{\pi}\,0.121\,.
\end{align}
Due to significant numerical cancellations for this particular value of $\eta$,
the radiative correction to the nuclear polarizability effect in the deuterium atom is quite small, $\sim 5$ Hz.
Therefore, it seems that the very good agreement for $r_d^2-r_p^2$ between the {\em electronic} and muonic determination \cite{pachucki:24a} is not accidental.

In summary, we have presented a comprehensive derivation of radiative corrections to the finite nuclear size effect in {\em electronic} and muonic atoms
We have verified the value for the leading radiative fns correction in Eq.  (\ref{34}) as obtained by Eides {\em et al.} \cite{eides:01} and corrected 
the result presented in Ref. \cite{pachucki:93b}. This is the leading radiative correction 
which should be included in the nuclear charge radii determination from the atomic spectroscopy.
The obtained numerical values for radiative recoil fns corrections are quite small in both
{\em electronic} and muonic atoms and thus can safely be neglected. 
Finally, we have obtained a closed formula for the radiative correction to the nuclear polarizability effect in {\em electronic} atoms.
In the case of deuterium,  this effect is  negligible due to accidental cancellations, and thus does not
affect the  $\delta r^2$ determination from the H-D isotope shift.
For muonic atoms, the evp correction to the nuclear polarizability effect
dominates and has already been accounted for in Ref. \cite{pachucki:24a},
while the muon self-energy and muon vacuum polarization corrections 
to the nuclear polarizability shift should be negligible.

\appendix

\section{Angle integration}
Let $A$ denote an average over the three-dimensional sphere in the Euclidean space,
\begin{align}
A[f] \equiv \int \frac{d\,\Omega_q}{2\,\pi^2}\,f(q,q_0) = \frac{2}{\pi}\int_0^\pi d\phi\,(\sin\phi)^2\, f\big(q,q\,\cos\phi\big)\,,
\end{align}
then
\begin{align}
A\biggl[q_0^2\biggr] =&\ \frac{q^2}{4}\,, \\
A\biggl[\frac{1}{q^4+4\,m^2\,q_0^2}\biggr] =&\  \frac{2}{q^4}\,\frac{1}{1 + \sqrt{1 + a^2}}\,,\\
A\biggl[\frac{1}{q_0^2}\biggr] =&\ -\frac{2}{q^2}\,,
\end{align}
where $a=2\,m/q$. In the second formula we assumed a symmetric integration around the pole at $q^0=0$, 
as denoted by subscript $s$ in Eq. (\ref{14}).

\begin{widetext}
\section{Master integral}
The master integral $J$ is defined as
\begin{align}
J(-q^2,q^0) =&\ -\int \frac{d^4 k}{\pi^2\,i}\;\frac{1}{k^2}\,\frac{1}{(t-k)^2-1}\;\frac{1}{(p-k)^2-1}
\nonumber \\ =&\
 \int_0^1 du \;\frac{1}{1-u(1-u)\,q^2 - u(1-p^2)}\;\ln
\left(\frac{1-u(1-u)q^2}{u(1-p^2)}\right), \label{B01}
\end{align}
where $p=q+t$ and $t=(1,\vec 0)$. 
We note, that in some of our former papers, i.e. Ref. \cite{pachucki:93a}, we used
a definition of $J$ which differs by a sign with Eq. (\ref{B01}).
The derivative of $J$ over the Euclidean angle, using Eq. (\ref{B01}), is
\begin{align}
\frac{\partial}{\partial \phi}\,\sin\phi\,J(q^2, i\,q\,\cos\phi) =&\ 
 -2\,i\,\sqrt{1 + \frac{4}{q^2}}\,\frac{\mathrm{arcsinh}\frac{q}{2}}{q - 2\,i\,\cos\phi} 
 + (i\,q + \cos\phi)\,\frac{\ln(q^2 - 2\,i\,q\,\cos\phi)}{q^2 - 2\,i\,q\,\cos\phi-1}. \label{B02}
\end{align}
From this one obtains another integral representation of $J$
\begin{align}
J(q^2, i\,q\,\cos\phi) =&\ \frac{1}{\sin\phi}\,\int_0^\phi d\phi'
\biggl[
 -2\,i\,\sqrt{1 + \frac{4}{q^2}}\,\frac{\mathrm{arcsinh}\frac{q}{2}}{q - 2\,i\,\cos\phi'} 
 + (i\,q + \cos\phi')\,\frac{\ln(q^2 - 2\,i\,q\,\cos\phi')}{q^2 - 2\,i\,q\,\cos\phi'-1}. 
 \biggr]
\end{align}
The particular form at $q^0=0$ of the master integral $J(q^2) \equiv J(q^2,0)$ is
\begin{align}
J(q^2) =&\ \int_0^1 du \;\frac{1}{1-u^2\,q^2}\;\ln\left(\frac{1+u(1-u)q^2}{u\,q^2}\right) \nonumber 
\\
=&\  1 + \frac{5\,q^2}{18} + \frac{11\,q^4}{150}  
 -\bigg(1 + \frac{q^2}{3} + \frac{q^4}{5}\bigg)\,\ln(q^2) + o(q^{10}). \nonumber 
\\
=&\ 
\frac{2}{q^2}+ \frac{2}{9\,q^4}  
 +\bigg(\frac{1}{q^2}  - \frac{5}{3\,q^4}  \bigg)\,\ln(q^2) + o(q^{-10})
\end{align}
The derivative of $J(q^2)$ satisfies
\begin{align}
\frac{\partial}{\partial q}  \big[ q J(q^2)\big] =&\ -\frac{4}{q^2}\,\frac{\mathrm{arcsinh}\frac{q}{2}}{\sqrt{1+\frac{4}{q^2}}} - \frac{\ln q^2}{1 - q^2}\,,
\end{align}
and from this one obtains another integral representation of $J$
\begin{align}
J(q^2) =&\ \frac{1}{q}\int_q^\infty dq'\,\biggl[
 \frac{4}{q'^2}\,\frac{\mathrm{arcsinh}\frac{q'}{2}}{\sqrt{1+\frac{4}{q'^2}}} + \frac{\ln q'^2}{1 - q'^2} \biggr].
 \end{align}
The imaginary part of $J$ in the sense of Eq. (\ref{31}) is
\begin{align}
J^A(q^2) =&\ -\frac{1}{q}\,\bigg[\arctan(q)-\Theta(q-2)\, \arccos\bigg(\frac{2}{q}\bigg)\bigg] \,.
\end{align}

\section{Two-photon amplitudes}
The electron (muon) self-energy $T_\mathrm{se}^{\mu\nu}$ correction to $T^{\mu\nu}$
can be represented as
\begin{align}
T_\mathrm{se}^{\mu\nu}(-q^2,q^0) =&\ \mbox{\rm Tr}\left[\gamma^\mu\left(
\Lambda(t,p,t) + 2\,\Gamma(t,p)\,\frac{1}{\not\!p-m} + 
\Sigma(p)\frac{1}{(\not\!p - m)^2}\right)\gamma^\nu
\frac{(\gamma_0+I)}{4}\right]\,,
\end{align}
where $\Sigma$ is the free electron self-energy, $\Gamma$ a vertex function, $\Lambda$ a double vertex function,
$p = t+q$, and $t=(1,\vec 0)$.
Because the symmetric tensor $T^{\mu\nu}_\mathrm{sym} = T_\mathrm{se}^{\mu\nu}(-q^2,q^0) 
+ T_\mathrm{se}^{\mu\nu}(-q^2,-q^0)$
satisfies $q_\mu\,T^{\mu\nu}_\mathrm{sym} = 0$, 
it can be expressed in terms of two independent functions $T^{00}_\mathrm{sym}$ and $ T^\mu_{\mathrm{sym},\,\mu}$,
where \cite{pachucki:95b}
\begin{align}
T_\mathrm{se}^{00}(-q^2,q^0) =&\ \frac{\alpha}{4\,\pi\,(1-p^2)}\,\biggl[ 
 (p^2-5)\,\biggl(\frac{(p^2-1)^2}{2\,q^2}  -p^2 -1 + \frac{q^2}{2}\biggr)
+ J\,\Big(3 + 10\,p^2 + 3\,p^4 - 6\,(1 + p^2)\,q^2 + 2\,q^4\Big) 
 \nonumber \\ &\
+ \frac{\arcsin\frac{q}{2}}{\sqrt{\frac{4}{q^2}-1}}\,\bigg(
41 + 16\,p^2 - p^4 - \frac{2\,(p^2-5)\,(p^2-1)^2}{q^4} + \frac{4\,(-9 - 16\,p^2 + p^4)}{q^2} + (p^2-9)\,q^2\bigg) 
\nonumber \\ &\
+ \frac{\ln(1 - p^2)}{2\,p^2}\,\Big(-1 + 17\,p^2 + 17\,p^4 - p^6 + (1 - 10\,p^2 + p^4)\,q^2\Big)
 \biggr]\,, \\
 T^\mu_{\mathrm{se},\,\mu}(-q^2,q^0)  =&\ \frac{\alpha}{4\,\pi\,(1-p^2)}\,\biggl[ 
 17 - 24\,p^2 - p^4 + (3 - 7\,p^2)\,q^2 
 + J\,\Big(6 + 12\,p^2 - 2\,p^4 + 4\,(p^2-1)\,q^2 - 4\,q^4\Big) 
 \nonumber \\ &\
 + \frac{\arcsin\frac{q}{2}}{\sqrt{\frac{4}{q^2}-1}}\,\Big(-32 + \frac{24\,(p^2-5)}{q^2} + 20\,q^2\Big)
 \nonumber \\ &\ 
 + \frac{(\ln(1 - p^2)+p^2)}{p^4}\,\Big(1 - p^2 + 15\,p^4 + p^6 + (-1 + 2\,p^2 + 7\,p^4)\,q^2\Big)
  \biggr]\,,
 \end{align}
where $p^2 = 1+q^2 + 2\,q^0$.
In the case $q^0 =0$  it is convenient to change to the Euclidean metric $q^2\rightarrow-q^2$.
With analytic continuation
\begin{align}
\frac{\arcsin\frac{q}{2}}{\sqrt{\frac{4}{q^2}-1}} \stackrel{q\rightarrow I\,q}{\rightarrow}&\ - \frac{\mathrm{arcsinh}\frac{q}{2}}{\sqrt{1+\frac{4}{q^2}}}\,,
\end{align}
one obtains
 \begin{align}
 T_\mathrm{se}^{00}(q^2,0) =&\ \frac{\alpha}{4\,\pi} \bigg[
 2 + \frac{8}{q^2} + \bigg(-4 + \frac{16}{q^2} - q^2\bigg)\,J + 
 8\,\bigg(1 - \frac{12}{q^4} - \frac{1}{q^2}\bigg)\,\frac{\mathrm{arcsinh}\big(\frac{q}{2}\big)}{\sqrt{1 + \frac{4}{q^2}}} 
 + \bigg(-20 + \frac{16}{q^2} + 3\,q^2\bigg)\,\frac{\ln(q^2)}{1 - q^2}\bigg] \nonumber 
 \\ =&\ \frac{\alpha}{4\,\pi} \frac{8}{9}\,\Big[5 - 6\,\ln(q^2)\Big] + o(q^2) \nonumber 
 \\ =&\ \frac{\alpha}{4\,\pi}\frac{2}{9\,q^2}\,\big[35 + 12\,\ln(q^2)\big] + o(q^{-4}). \label{C05}
 \end{align}
The imaginary part of $T^{00}$  in the sense of Eq. (\ref{31}) is
\begin{align}
T_\mathrm{se}^{00A}(q^2) =&\ \frac{\alpha}{\pi}\,\biggl[-\frac{3}{4} - \frac{4}{q^2} - \frac{1}{4\,(1 + q^2)} + J^A\,\biggl(-1 - \frac{4}{q^2} + \frac{q^2}{4}\biggr) 
+ \frac{\Theta(q-2)}{\sqrt{1-\frac{4}{q^2}}}\,\biggl(1+\frac{1}{q^2}-\frac{12}{q^4}\biggr)\biggr]. \label{C08}
\end{align}

\end{widetext}

\end{document}